\documentclass[letterpaper]{article} 
\usepackage{aaai24}  
\usepackage{times}  
\usepackage{helvet}  
\usepackage{courier}  
\usepackage[hyphens]{url}  
\usepackage{graphicx} 
\usepackage{natbib}  
\usepackage{caption} 
\usepackage{algorithm}
\usepackage{algorithmic}
\usepackage{multirow}
\usepackage{bm}
\usepackage{newfloat}
\usepackage{listings}
\DeclareCaptionStyle{ruled}{labelfont=normalfont,labelsep=colon,strut=off} 
\floatstyle{ruled}
\newfloat{listing}{tb}{lst}{}
\floatname{listing}{Listing}
\title{Generative Multi-Modal Knowledge Retrieval with Large Language Models}

\author {
    Xinwei Long\thanks{The work was done when Xinwei Long worked as intern at Pattern
Recognition Center, WeChat AI, Tencent Inc, China.}\textsuperscript{\rm 1},
    Jiali Zeng\textsuperscript{\rm 2},
    Fandong Meng\textsuperscript{\rm 2},
    Zhiyuan Ma\textsuperscript{\rm 1},
    Kaiyan Zhang\textsuperscript{\rm 1},\\
    Bowen Zhou\thanks{Corresponding author}\textsuperscript{\rm 1 },
    Jie Zhou\textsuperscript{\rm 2}
}
\affiliations {
    \textsuperscript{\rm 1}Department of Electronic Engineering, Tsinghua University, Beijing, China\\
    \textsuperscript{\rm 2}Pattern Recognition Center, WeChat AI, Tencent Inc, China\\
    longxw22@mails.tsinghua.edu.cn, zhoubowen@tsinghua.edu.cn
}

\usepackage{bibentry}

\begin{document}

\maketitle

\begin{abstract}
Knowledge retrieval with multi-modal queries plays a crucial role in supporting knowledge-intensive multi-modal applications. However, existing methods face challenges in terms of their effectiveness and training efficiency, especially when it comes to training and integrating multiple retrievers to handle multi-modal queries. 
In this paper, we propose an innovative end-to-end generative framework for multi-modal knowledge retrieval. 
Our framework takes advantage of the fact that large language models (LLMs) can effectively serve as virtual knowledge bases, even when trained with limited data. We retrieve knowledge via a two-step process: 1) generating knowledge clues related to the queries, and 2) obtaining the relevant
document by searching databases using the knowledge clue.
In particular, we first introduce an object-aware prefix-tuning technique to guide multi-grained visual learning.
Then, we align multi-grained visual features into the textual feature space of the LLM, employing the LLM to capture cross-modal interactions. 
Subsequently, we construct instruction data with a unified format for model training.
Finally, we propose the knowledge-guided generation strategy to impose prior constraints in the decoding steps, thereby promoting the generation of distinctive knowledge clues. 
Through experiments conducted on three benchmarks, we demonstrate significant improvements ranging from 3.0\% to 14.6\% across all evaluation metrics when compared to strong baselines.
\end{abstract}

\section{Introduction}

\begin{figure}[t]
\centering
\includegraphics[width=1.0\columnwidth]{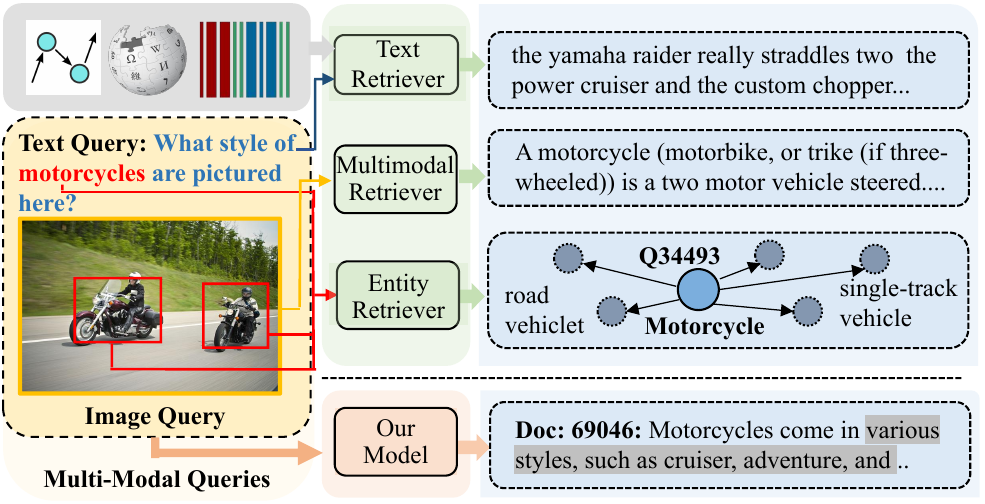} 
\caption{Multi-Modal Knowledge Retrieval.
Prior studies use multiple retrievers for separate purposes, while we retrieve knowledge through an end-to-end generative model.}
\label{fig1}
\end{figure}

Knowledge Retrieval (KR) is crucial in supporting knowledge-intensive multi-modal applications, such as visual question answering (VQA)~\cite{ma2023hybridprompt}, multi-modal entity linking~\cite{10.1007/978-3-031-17189-5_24} and multi-modal dialogue~\cite{ma2022unitranser}. 
In these applications, the information available within the multi-modal contexts may be insufficient, necessitating the acquisition of external knowledge.
As illustrated in Fig.~\ref{fig1}, knowledge retrievers offer key evidence to assist VQA systems in identifying the motorcycle's style as a ``chopper". 
In recent years, information retrieval~\cite{DBLP:journals/fcsc/ChenCDDGHLLLLLM21,DBLP:journals/tois/WuZGFC23,DBLP:conf/kdd/Tang0GCZWYC23} has achieved remarkable success. However, challenges still persist in terms of effectiveness and training efficiency when applying these methods to multi-modal scenes.
Existing methods~\cite{DBLP:conf/emnlp/LuoZBB21,gao2022thousand} handle multi-modal queries by utilizing individual text-to-text and image-to-text retrievers, which struggle to capture cross-modal interactions and require abundant data to train each module in the pipelines.
The question arises:
\textit{Can we develop a retriever that effectively handles multi-modal queries while avoiding the redundant pipeline?}

 Recently, there has been a promising development in the field of Ad-hoc retrieval called generative retrieval~\cite{DBLP:conf/nips/WangHWMWCXCZL0022,DBLP:conf/nips/BevilacquaOLY0P22}. 
 This approach aims to simplify the retrieval pipeline by generating relevant document identifiers instead of retrieving them from a large-scale corpus.
 Instead of retrieving actual documents, these methods directly generate identifiers such as document titles or URLs that are relevant to the query.
 
Nevertheless, these generative retrieval methods have not been applied to multi-modal knowledge retrieval for two reasons.
 Firstly, knowledge-aware documents, which contain information from multiple aspects, cannot be effectively represented by static identifiers, such as numeric IDs~\cite{DBLP:conf/nips/Tay00NBM000GSCM22} and titles~\cite{10.1145/3511808.3557271}.
 This is because queries from different modalities attend to different aspects of documents.
 For example, as depicted in Fig.~\ref{fig1}, the text query attends to keywords that are present in both query and document (e.g. ``motorcycles''), whereas the image query concentrates on descriptive words about specific visual elements.
 Secondly, 
 the identifiers~\cite{DBLP:conf/nips/Tay00NBM000GSCM22} require additional memory steps and are inefficient when dealing with large-scale corpora. 
 This approach proves to be training-inefficient and struggles to perform well when encountering unseen knowledge, highlighting its lack of generalization capabilities.
To address these challenges, we propose a generative framework for multi-modal knowledge retrieval, briefly denoted as \textbf{GeMKR}. 
This framework leverages the LLMs, LLaMA, as its core model, based on the premise that LLMs can effectively function as virtual knowledge bases (KB)~\cite{pan2023unifying} even then fine-tuning with limited data.
 In GeMKR, we abandon the traditional pipeline that calculates the similarity between queries and knowledge.
Instead, we retrieve knowledge via a two-step process: 1) generating knowledge clues related to the queries, and 2) obtaining the relevant document by searching databases using the knowledge clue.
Please note that only the first step requires neural computation, while the second step is a definitive and efficient database operation.
Here, knowledge clues are defined as any subsequences within a document that appear exclusively in that particular document.
Unlike the one-to-one relationship between an identifier and a document, each knowledge clue in GeMKR uniquely corresponds to a knowledge-aware document in the knowledge base, while each document can be mapped by multiple knowledge clues.
We carry out primary experiments on three benchmarks, with knowledge base sizes ranging from 112,724 to 21,015,324.
The experimental results show significant improvements of 3.0-14.6\% across all metrics compared to strong baselines.
Notably, GeMKR achieves improvements of 14.6\% and 8.9\% in P@5 and R@5 respectively, when retrieving information from a knowledge base comprising 21 million documents. This outcome illustrates our model's capacity to generalize well to large-scale knowledge sources.

\section{Related Work}
\subsection{Multi-Modal Knowledge Retrieval}
Knowledge-intensive multi-modal tasks require extensive knowledge access due to the insufficiency of vital information within their contexts. 
Existing methods ensemble various types of retrievers to acquire world knowledge. 
Representative retrieval methods include BM25~\cite{10.1561/1500000019} and DPR~\cite{karpukhin-etal-2020-dense} for text retrieval, CLIP~\cite{radford2021learning} for image-to-text retrieval, and GENER~\cite{decao2021autoregressive} for entity retrieval. However, the simple integration of multiple retrievers for individual purposes is inadequate for knowledge-intensive multi-modal tasks due to the following reasons. 
Firstly, it is important to consider the interaction between visual and textual queries in order to understand the relationships between visual objects and textual entities. Secondly, the pipeline involves the integration of various external tools, resulting in inconvenient usage. 

In contrast to the integration of multiple traditional retrievers, some studies have proposed new methods and benchmarks to facilitate research on this task. ~\cite{luo2021weakly} constructs a knowledge retrieval dataset using the OKVQA benchmark~\cite{okvqa}. 
This dataset necessitates the retrieval of relevant evidence, using both the question and image as queries. 
This dataset comprises a small KB of 112K knowledge records and a large KB consisting of 21M records. This poses a tough challenge in obtaining accurate knowledge from such an extensive KB. \cite{DBLP:conf/acl/0003FGYB23} introduces a high-quality multi-modal knowledge retrieval dataset, imposing higher demands on cross-modal understanding. 
To jointly encode visual and textual queries, recent studies~\cite{DBLP:conf/acl/0003FGYB23} have explored training a single-stream vision-language model to obtain cross-modal representation. 
Due to differences between the modalities, they use millions of data to train their models in a contrastive framework. 
Despite achieving improvements over the above methods, multi-modal knowledge retrieval remains an under-explored task in terms of effectiveness and training efficiency.

\subsection{Generative Retrieval}
Recently, some studies~\cite{DBLP:conf/nips/WangHWMWCXCZL0022,DBLP:conf/nips/Tay00NBM000GSCM22,DBLP:conf/acl/0001YWWL23,zhou-etal-2023-enhancing-generative} have explored retrieving documents through generative language models, e.g. BART. They simplify the pipelines of retrieval by directly generating identifiers of relevant documents for queries rather than retrieving from a large-scale corpus. 
\cite{DBLP:conf/nips/Tay00NBM000GSCM22} proposes DSI (Differentiable Search Index) framework that builds the search index in Transformer memories rather than in databases. It assigns each document a numeric ID as the identifier, which requires extra memory steps and is inefficient and ineffective in the large-scale corpus.
Instead of numeric IDs, \cite{10.1145/3511808.3557271,DBLP:conf/sigir/ChenZGFC22} take Wikipedia titles as identifiers to integrate semantic information about documents into identifiers, whereas ~\cite{DBLP:conf/nips/BevilacquaOLY0P22,DBLP:conf/acl/0001YWWL23,DBLP:conf/sigir/Chen0GR0FC23} leverages n-grams in documents as identifiers and introduces an efficient FM-Index~\cite{DBLP:conf/focs/FerraginaM00} structure to guide the generation of identifiers.
However, 
each n-gram could correspond to several documents, as a short n-gram might appear in several contexts. To address these issues, the above methods generate numerous n-grams for each query and then re-rank these n-grams to obtain the final results.

\subsection{Large Language Model and Efficient Fine-tuning}
Early explorations leveraging the power of LLMs for information retrieval exist, such as using LLMs to understand queries~\cite{jagerman2023query}, generating training data for downstream retrieval~\cite{gao-etal-2023-precise}, and making decisions in re-ranking stages~\cite{ferraretto2023exaranker}, which demonstrate the potential of LLMs in the retrieval task.
Despite the strong capabilities of LLMs, their enormous parameters pose challenges for computational resources when fine-tuning for downstream tasks.
To mitigate these challenges, parameter-efficient fine-tuning methods~\cite{han2021ptr,2023:delta,zhang-etal-2023-crash} have been proposed which reduce costs by updating only a subset of parameters. Typical efficient methods include prompt tuning~\cite{ding2021openprompt}, prefix tuning~\cite{DBLP:conf/acl/LiL20,DBLP:conf/iclr/YangL22a}, adapter methods~\cite{DBLP:journals/corr/abs-2303-16199,DBLP:conf/acl/DiaoXXWZ23}, and the low-rank (LoRA) method~\cite{hu2021lora}.

\section{Methodology}
\begin{figure}[t]
\centering
\includegraphics[width=1.0\columnwidth]{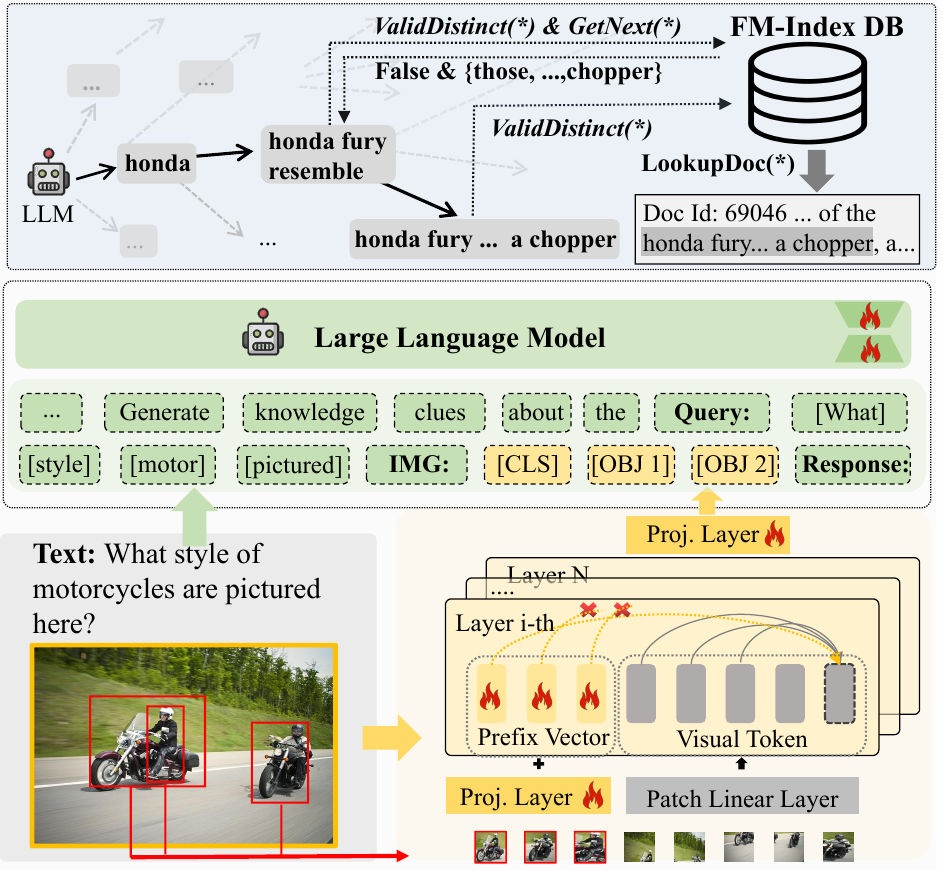} 
\caption{Overall architecture of our GeMKR.}
\label{fig2}
\end{figure}

We propose GeMKR, an end-to-end generative framework for multi-modal knowledge retrieval.
GeMKR consists of three components, as depicted in Fig.~\ref{fig2}: {\bf Object-aware prefix-tuning} for fine-tuning the visual backbone, {\bf Multi-Modal Alignment} using LLMs to capture cross-modal interactions, and {\bf Knowledge-guided Constraint Decoding} for generating informative knowledge clues.
\subsection{Problem Formulation}
Formally, let $\mathcal{D}=\{D_1, D_2,...,D_i\}_{i=1}^{N}$ denotes a knowledge base used for the multi-modal knowledge retrieval task. As a generative retrieval model, our goal is to generate the relevant knowledge clues $\{C\}$, which can be definitively mapped to documents in $\mathcal{D}$.
Our model takes the multi-modal query ${Q}=\{T, V\}$ as input and generate the relevant knowledge clue with an auto-regressive score, as Eq.~\ref{eq1}.
\begin{equation}
  {\rm score}(C,Q) = \mathcal{F}_{\Theta}(C|Q) = \prod_{j=1}\mathcal{F}_{\Theta}(c_j|\bm {c}_{<j},Q) . 
  \label{eq1}
\end{equation}
where $\mathcal{F}_{\Theta}$ is the generative retriever with parameters $\bm \Theta$, and $c_j$ is the $j_{th}$ token of the knowledge clue. During inference, the model employs a constrained strategy to guide the decoder in generating content within a limited token space at each step, which ensures that each knowledge clue $C_j$ can be definitively mapped to a document $D_i$ as Eq.~\ref{eq2}.
\begin{equation}
  \varphi: C_j \to D_i, {\rm where}\  D_i \in \mathcal{D} .
  \label{eq2}
\end{equation}
Finally, we sort the document set $\{D_i\}$ based on auto-regressive scores $\rm{score}(C_i,Q)$ to obtain the final retrieval results. 

\subsection{Object-aware Prefix Tuning}
To efficiently fine-tune the visual backbone, we present the Object-aware Prefix Tuning method that explicitly guides the visual understanding using objects (i.e., visual entities) as the learnable prefix. 
As shown in the bottom of Fig.~\ref{fig2}, we utilize the CLIP model with $N$ Transformer layers as our visual backbone, and we feed two groups of features, $\bm X_P$ and $\bm X_I$, into CLIP using a prefix tuning approach as Eq.~\ref{eq3}.
Here, $\bm X_P$ denotes the learnable prefix prompts, which are mixed with fine-grained visual object information, while $\bm X_I$ represents the embeddings of each visual token encoded by the patch embedding layer. 
To mitigate catastrophic forgetting of the visual backbone, 
we freeze the parameters $\bm \Theta_v$ of the visual backbone and only make the prefix prompts $\bm X_P$ learnable.
\begin{equation}
  H_v= \mathcal{F}_{\rm visual}({\bm X_P,\bm X_I;\bm \Theta_v}). 
  \label{eq3}
\end{equation}

For the preparation of $\bm X_P$, we first randomly initialize the $N$ prefix prompts $\bm X_{Pr}$ for $N$ layers, with the parameter matrix in a dimension of $l \times d$, where $l$ denotes the length of the prefix and $d$ is the visual dimension. Note that the prefix prompts are not shared across layers.
To obtain object features, we crop the objects from images and transform them into a fixed resolution. Next, we extract their features as $\bm X_R \in \bm \Re^{r \times d}$ using a frozen visual encoder, CLIP. Lastly, we feed $\bm X_R$ into a learnable projection layer as $\bm H_R = {\rm Linear}(\bm X_R)$ and pad them with zero vectors to maintain the same dimension as the prefix.
Taking the $\rm i^{th}$ layer in the visual backbone as an example, we denote the predefined prefix vector as $\bm X_{Pr}^i\in \bm \Re^{l\times d}$ and the visual features obtained from the $(i-1)^{th}$ layer is $\bm H_I^{i-1}\in \bm \Re^{l\times d}$, we acquire the object-aware prefix vector via simply addition as Eq.~\ref{eq4},
\begin{equation}
  \bm X_P^i= \bm X_{Pr}^i + \bm H_R .
  \label{eq4}
\end{equation}
After that, we concatenate $\bm X_P^i$ in front of $\bm X_I^i$, as $\bm X=[\bm X_P^i; \bm H_I^i]$, and feed them into the self-attention module. This method allows the fine-grained object features within $X_P^i$ to better guide the visual backbone during the adaptation process.

However, the distribution discrepancy between the object feature $\bm H_R$ and the immediate output $\bm H_I^i$ may lead to a significant loss at the early training stages, potentially disturbing the fine-tuning process.
Similar to ~\cite{DBLP:journals/corr/abs-2303-16199}, we introduce a dual-flow attention mechanism to address this issue by independently computing attention weights for prompt vectors and hidden states.
Specifically, We compute the query vector on $\bm H_I^i$, by applying ${\rm Linear_{Q_i}}(\bm H_I^i)$, while the key and value vectors are independently computed for $\bm X_P^i$ and $\bm H_I^i$ using the linear layers ${\rm Linear}_{K_i} (\cdot)$ and ${\rm Linear}_{V_i} (\cdot)$, denoted as $\bm K^i_p$, $\bm K^i_h$, $\bm V^i_p$, and $\bm V^i_h$. We follow the vanilla attention module, as Eq.~\ref{eq5}, but individually apply attention maps for two components and multiply value vectors to obtain outputs, as Eq~\ref{eq6},
\begin{equation}
  {\rm Att}(\bm Q, \bm K, \bm V)= {\rm Softmax}(\frac{\bm Q^i(\bm K^i)^T}{\sqrt{d}}) \bm V ,
  \label{eq5}
\end{equation}
\begin{equation}
  \bm H^i_O = {\rm Att}(\bm Q^i_h, \bm K^i_p, \bm V^i_p) *\sigma + {\rm Att}(\bm Q^i_h, \bm K^i_h, \bm V^i_h) .
  \label{eq6}
\end{equation}
Here, $\sigma$ denotes a gate function to control the information flow from the prompt vectors to visual tokens. Dual-flow attention mechanism effectively alleviates the influence of uncertainty from the prefix, while keeping the ability of the visual backbone to obtain high-quality representation, thereby making a more stable fine-tuning process. After $N$ Transformer Layers, we could obtain the final representation $\bm H_v$ from the visual backbone, where $\bm H_v$ contains a representation of the $\rm [CLS]$ token, denoted as $\bm h_{\rm cls}$ and representations of other visual tokens.

\subsection{Multi-Modal Alignment}
To effectively integrate visual features into pre-trained LLM, we employ the simple projection scheme as illustrated in the middle part of Fig.~\ref{fig2}, which demonstrates effectiveness in other vision-language studies~\cite{DBLP:journals/corr/abs-2304-10592}.
Specifically,
we utilize the $\rm [CLS]$ token $\bm h_{\rm cls}$ to represent the image at a holistic level since the long sequence of visual tokens would disturb the linguistic knowledge in LLMs. Besides, we also leverage object features $\bm H_R$ as features to integrate object-level visual information. 
Then, a simple linear layer is applied to map the visual representation to the text embedding spaces as Eq.~\ref{eq8}, 
\begin{equation}
  \bm H_v^{'}= {\rm Linear}([\bm h_{\rm cls};\bm H_R]) .
  \label{eq8}
\end{equation}
After that, we utilize LLaMA as our textual backbone, with text embedding and the visual representation $\bm H_v^{'}$ as inputs. Based on the multi-modal input, the LLM can predict the next token step by step.

\subsection{Instruction Tuning}
\subsubsection{Supervised Text Sampling.} 
During training, knowledge clues are not explicitly provided in this task. 
Instead, the benchmarks offer a set of relevant documents for each query. 
However, these documents tend to be excessively long, comprising redundant information, whereas knowledge clues are ideally concise text snippets directly pertinent to queries. 

To address this issue, we first split the positive document into individual sentences and evaluate each sentence's relevance by counting the number of keyword hits between the sentence and the query,
where the higher the hit rate, the more relevant information the sentence contains. 
Despite selecting the sentences $S$ with the most keyword hits, they are still not ideal supervised texts, since these sentences have varying lengths. Therefore, we calculate the count of hits $c_{w_i}$ for each span $w_{i:i+n}$ as the relevant score $\pi_{w_i} = \frac{c_{w_i}}{c_{w_i} + \rho}$ and normalize the scores of all the spans in the sentence through a softmax function, where $\rho$ is the smoothing factor. We sample $m$ start positions of snippets according to the normalized distribution and cut out $l$ tokens from the start positions to obtain $m$ knowledge clues with the same length.

\subsubsection{Instruction Data Construction.} Firstly, we create the instruction template in a unified format with predefined slots for filling the multi-modal queries. 
As shown in Fig~\ref{fig2}, the template contains a task description, an instruction text derived from the textual query, and several predefined slots for visual features. The instruction data is used to train the model, prompting the prediction of tokens after the ``response:" token, and thus only the predicted tokens are used to compute the loss.

\subsubsection{Model Training.} We perform instruction tuning for the whole model on the predicted tokens. We freeze the parameters of LLaMA and use the low-rank adaptation (LoRA) method for efficient adaptation. We adopt the common auto-regressive training objective, as Eq~\ref{eq10},
\begin{equation}
  \mathcal{L}_{\rm gen}= -\sum_{i=1}^l{
  \rm log} \mathcal{F}_{\Theta}(c_i|\bm {c}_{<i};T;V;\bm \Theta) .
  \label{eq10}
\end{equation}
where $\Theta$ denotes unfrozen parameters.



\subsection{Knowledge-guided Constraint Decoding}
During inference, our model applies the knowledge-guided constraint decoding strategy to guide the decoder in searching within a limited token space at each step, so as to generate a valid knowledge clue that can be mapped to one and only one document within the knowledge base. 
To facilitate efficient search from the KB, we introduce the FM-Index database~\cite{DBLP:conf/focs/FerraginaM00} for its storage. The FM-Index offers three interfaces: \textit{GetNext}, \textit{ValidDistinct}, and \textit{LookupDoc},  enabling efficient lookup from a large-scale corpus at the millisecond level.
In each generation step, our model employs the previously generated tokens as prefix conditions to invoke the \textit{GetNext} interface. 
Subsequently, the interface searches for the strings that match this prefix, obtaining the succeeding token as the next allowable token.
Lastly, the model constructs a mask matrix derived from the set of allowable tokens, wherein tokens in the set are assigned a value of 1, and others are set to 0.
This matrix is employed to modify the predicted distribution, ensuring that the decoded knowledge clue appears at least once in the knowledge base.




\begin{table*}[!th]
\centering
\begin{tabular}{lccccccccc}
\hline\hline
\multirow{2}{*}{Model}                                    & \multicolumn{3}{c}{OKVQA-GS112K} & \multicolumn{3}{c}{OKVQA-WK21M} & \multicolumn{3}{c}{ReMuQ} \\ \cline{2-10} & P@5   & R@5       & R@10     & P@5       & R@5      & R@10     & P@1     & R@5    & R@10   \\ \hline
BM25~\cite{DBLP:journals/ftir/RobertsonZ09} $\triangle$   & 27.5      & 51.4      & 63.0     & 27.9      & 50.2     & 60.9     & 5.6     & 8.8    & 10.8   \\
DPR~\cite{DBLP:conf/emnlp/KarpukhinOMLWEC20} $\triangle$ & 27.7      & 55.6      & 66.4     & 28.1      & 59.4     & 71.1     & 35.8    & 43.4   & 48.8   \\
CorpusBrain~\cite{10.1145/3511808.3557271} $\star$          &    28.2   &    58.6    &    66.9    &     -    &     -    &     -   &     -    &   -    &   -    \\
SEAL~\cite{DBLP:conf/nips/BevilacquaOLY0P22}   &    30.4    &   62.9    &  73.9   &    -     &    -    &    -    &   56.7     &   66.4  & 74.1      \\ \hline
CLIP~\cite{DBLP:conf/icml/RadfordKHRGASAM21} $\triangle$  & 11.1      & 34.5      & 50.5     & 9.7       & 29.8     & 43.0     & 19.4    & 40.2   & 49.3   \\ 
VRR~\cite{DBLP:conf/emnlp/LuoZBB21} $\triangle\circ$         & 39.4      & 71.5      & 81.5     & -         & -        & -        & -       & -      & -      \\
ReViz\cite{DBLP:conf/acl/0003FGYB23} $\triangle$         & 34.5      & 66.1      & 77.8     & 30.1      & 60.9     & 72.2     & 49.1    & 62.4   & 71.6   \\
ReViz-ICT\cite{DBLP:conf/acl/0003FGYB23} $\triangle\star$  & \underline{41.7}      & \underline{73.4}      & \underline{83.2}     & \underline{31.4}      & \underline{61.9}     & \underline{72.6}     & \underline{62.1}    & \underline{76.2}   & \underline{83.3}   \\ \hline
GeMKR (Our Model)                                         & \textbf{49.1}      & \textbf{78.6}      & \textbf{86.2}     & \textbf{46.0}      & \textbf{70.8}     & \textbf{79.1}     & \textbf{75.2}    & \textbf{90.3}   & \textbf{92.7}   \\ \hline\hline
\end{tabular}
\caption{Results on the benchmarks of multi-modal knowledge retrieval, where $\star$ represents the method uses external data, $\circ$ indicates the method ensembles several retrievers, $\triangle$ means results are reported in \cite{DBLP:conf/acl/0003FGYB23}. }
\label{mainexp}
\end{table*}
 
To make the generated knowledge clues more distinctive, we force the model to generate at least $l_{min}$ tokens. Then, we use the \textit{ValidDistinct} interface to validate whether the generated tokens can be uniquely mapped to a knowledge record in the KB. If the return is ``True'', the generation process is stopped. 
Otherwise, the model continues to generate the next token and validate every step until the max length $l_{max}$ is reached or the returned value is ``True''. 
We add the penalty term in the decoding process to encourage short discriminable clues to be ranked in the front of the queue whereas long ambiguous clues are in the tail. 
Based on these strategies, most of the generated clues can correspond to a unique record in the knowledge base. We regard the knowledge clues corresponding to several documents as invalid outputs and drop them directly. We can obtain the whole knowledge document by using the \textit{LookupDoc} interface with a generated knowledge clue as input, which is a definitive and efficient operation.

\section{Experiments}
\subsection{Settings}
We conduct experiments on three benchmarks of multi-modal knowledge retrieval:  OKVQA-GS112K~\cite{DBLP:conf/emnlp/LuoZBB21}, OKVQA-WK21M~\cite{luo-etal-2023-end} and ReMuq~\cite{luo-etal-2023-end}, which are derived from the VQA task leveraging both the image and question as queries.
The dataset statistics can be found in Tab.~\ref{tab1}.


\begin{table}[t]
\centering
\begin{tabular}{ccccccc}
\hline\hline
Dataset      & Train/Val/ Test & KB size  \\ \hline
OKVQA-GS112K     & 8,062/896/5,046 & 112,724                  \\
OKVQA-WK21M    & 8,062/896/5,046 & 21,015,324               \\
ReMuq            & 7,576/842/3,609 & 195,837                   \\
\hline\hline
\end{tabular}
\caption{Dataset statistics.}
\label{tab1}
\end{table}

\subsubsection{Evaluation Metrics.} 
We strictly follow the settings of the original papers, using the corresponding metrics for each dataset. We evaluate model performance using Pseudo-relevance Precision@K (P@K) and Pseudo-relevance Recall@K (R@K). Specifically, we use R@5, and R@10 for all datasets. For ReMuQ, which has exactly one correct document per query, we use P@1. For the other datasets, we use P@5. Please refer to the formalized definition in their original paper~\cite{DBLP:conf/emnlp/LuoZBB21,DBLP:conf/acl/0003FGYB23}.

\subsubsection{Baselines.} 
We adopt several baseline methods for comparison: (1) BM25~\cite{DBLP:journals/ftir/RobertsonZ09} and DPR~\cite{DBLP:conf/emnlp/KarpukhinOMLWEC20} are classical document retrieval models. (2) CorpusBrain~\cite{10.1145/3511808.3557271} and SEAL~\cite{DBLP:conf/nips/BevilacquaOLY0P22} are advanced generative retrieval models. (3) CLIP~\cite{DBLP:conf/icml/RadfordKHRGASAM21} is a typical image-to-text retriever. 
(4) VRR ~\cite{DBLP:conf/emnlp/LuoZBB21} 
integrates three retrievers, including BM25, DPR, and LXMERT~\cite{tan2019lxmert}.
(5) ReViz and ReViz+ICT~\cite{DBLP:conf/acl/0003FGYB23} are multi-modal retrievers that are designed for this task. Note that we use the image caption model to obtain the textual description of images and feed the textual features to enhance the understanding of multi-modal contexts for textual baselines.




\subsubsection{Implementation Details}
In our main experiments, we use ViT-L/14 from pre-trained CLIP~\cite{DBLP:conf/icml/RadfordKHRGASAM21} as the image encoder and LLaMa-7b~\cite{touvron2023llama} as the text encoder. We use YOLOv7~\cite{wang2022yolov7} to obtain bounding boxes, keeping the top 5 most confident objects for images with excessive objects.
Our model is implemented by Pytorch and trained using a learning rate of 6e-5, the Adam optimizer with a warm-up strategy, and batches of 12 instruction data.
Training is performed on an NVIDIA A6000 48G GPU and completed within three hours.
Unlike other approaches, we train our model end-to-end without additional data. We construct instruction data from the original dataset as shown in Tab.~\ref{tab1}, and 
sample two knowledge clues for each positive document.
For inference, knowledge sources are indexed using the FM-Index~\cite{DBLP:conf/focs/FerraginaM00} technique and stored in the Sdls-lite\footnote{https://github.com/simongog/sdsl-lite} database for efficient storage and lookup. To generate distinct knowledge clues, we use constrained beam search to decode clues over 10-15 timesteps with 20 beams and 4 beam groups. \footnote{The code will be released in this repository. https://github.com/xinwei666/MMGenerativeIR}


\subsection{Main Results}


As shown in Tab~\ref{mainexp}, we conduct a comparative analysis of our model against baseline approaches across three benchmarks, varying in KB sizes from 112K to 21M. 
Evidently, our proposed approach consistently outperforms the leading state-of-the-art baselines across all evaluated metrics.
In particular, the improvements, measured by P@K, surpass a minimum of 13.1\% on the ReMuQ and OKVQA-WK21M datasets, demonstrating our model's capacity to retrieve more precise knowledge compared to alternative baselines.
Besides, ReViz-ICT, which utilizes a single-stream query encoder to capture cross-modal interactions, consistently achieves superior performance among other baselines. The lower performance of other baseline models,
emphasizes the importance of cross-modal interaction in this task. 
However, training a multi-modal query encoder necessitates a large amount of multi-modal data, which is both resource-intensive and data-inefficient. 
In contrast, our model only requires 20K instruction data for lightly fine-tuning with only 14M parameters of the total 7.3B.
What's more, we observe that the textual generative baseline SEAL produces promising results when employing image captions as visual features. This observation indicates the effectiveness of generative models in knowledge retrieval tasks.

While ReViz-ICT demonstrates good performance in retrieving from smaller knowledge bases, the improvement is less evident when applied to the larger-scale knowledge base. 
Conversely, our model exhibits superior performance,
surpassing ReViz-ICT by a margin of at least 6.4\%, when retrieving information from a knowledge base comprising 21 million entries.
This outcome demonstrates that our model can generalize well to varying scales of knowledge sources.

In a nutshell, our model is well-suited for multi-modal knowledge retrieval, supported by two potential reasons. Firstly, our model adeptly aligns visual representations with LLMs, thereby enhancing its capability to deeply understand multi-modal queries.
 Secondly, our model introduces a constrained beam search guided by knowledge bases. This approach takes advantage of the knowledge potential of LLMs while imposing constraints to mitigate unreliable outputs.

\section{Analysis}

\subsection{Ablation Study}
\begin{table}[t]
\centering
\begin{tabular}{lcccc}
\hline\hline
Delete Module       & P@5  & R@5      & R@10 \\ \hline
Full Model          & 49.1 & 78.6     & 86.2 \\ \hline
\ w/o Obj. Feature        & 46.9 & 76.7     & 85.3 \\
\ w/o Dual-flow Att   & 47.5 & 76.5     & 84.5 \\
\ w/o Obj. Prefix              & 46.5 & 76.5     & 84.1 \\ \hline
\ w/o Lora               & 44.5 & 75.3 & 84.2 \\
\ w/o (Obj. Prefix \& Lora)     & 39.5 & 72.2     & 82.1 \\ \hline
\ w/o Visual Queries     & 40.4 & 69.8    & 79.3 \\ \hline\hline
\end{tabular}
\caption{Ablation studies on the OKVQA-GS112K .}
\label{ablation}
\end{table}
In this section, we conduct a series of ablation studies from the bottom to the top layer by deleting each module respectively. Results are in Tab~\ref{ablation}.

In the ablation study of the Object-aware Prefix Tuning (Obj. Prefix) module, we initiate the evaluation by omitting the object feature, resulting in an evident drop of 2.2\% and 1.9\% in P@5 and R@5 respectively. 
This observation indicates the importance of object features in multi-modal knowledge retrieval.
Next, we substitute the dual-flow attention mechanism (as Eq~\ref{eq6}) with the vanilla attention, leading to a performance decrease of 1.5\% to 2.1\%. 
Lastly, upon complete removal of the prefix tuning module, there is a significant decline of at least 2\% across all metrics, thus demonstrating the effectiveness of object-aware prefix-tuning in integrating multi-grained visual features.

Furthermore, we explore the effectiveness of each operation on the LLMs. 
We first freeze all parameters of the LLM (i.e. removing the Lora adaptation), resulting in a performance decrease. 
Additionally, when both the Object-aware Prefix-tuning and Lora adaptation are removed (only updating the parameters of the projection layers), the results exhibit a sharp decline of over 10\%, falling below even the baseline performance levels. 
This outcome demonstrates
the essential role of the LLM, which operates as a virtual knowledge base for generating precise knowledge clues. 
Finally, we remove the visual module and use textual queries and image captions as inputs. 
The model's performance decreases further, highlighting that image caption models cannot replace the role of visual modules in multi-modal tasks. 
Nevertheless, the performance is still better than the best textual baseline SEAL, which shows the effectiveness of other designs in our model. 


\subsection{Effect of Model Sizes}
\begin{figure}[t]
\centering
\includegraphics[width=0.99\columnwidth]{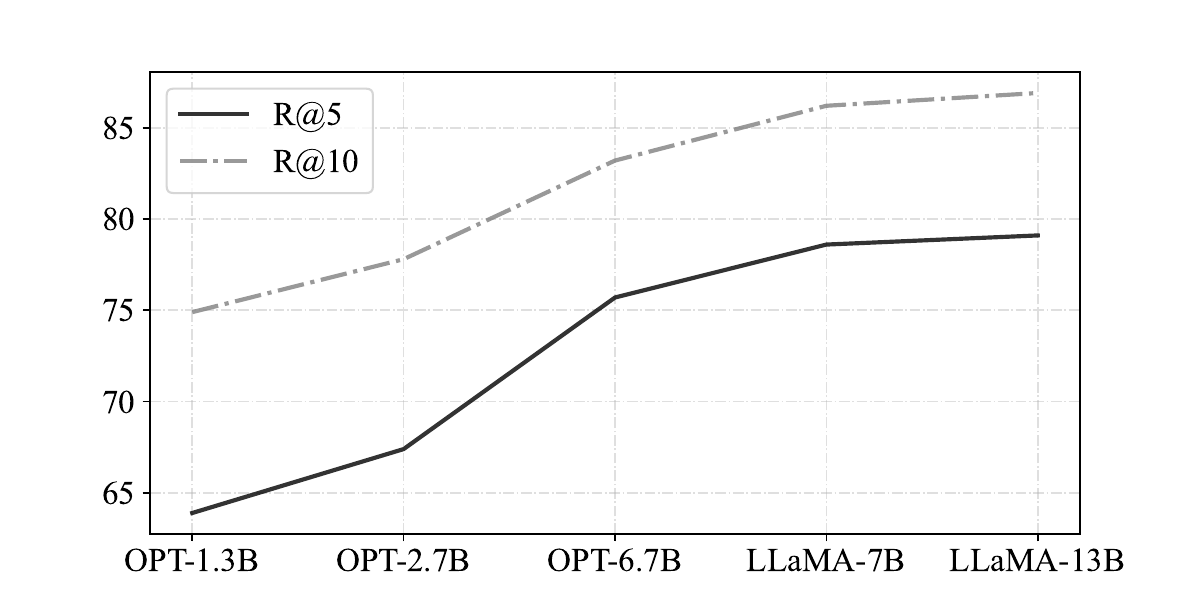} 
\caption{Results of scaling up the LLMs}
\label{modelscale}
\end{figure}

Additionally, we utilize diverse LLMs~\cite{zhang2022opt,touvron2023llama} at varying scales (i.e., 1.3B, 2.7B, 6.7B, 7B, and 13B) to examine the impact of the LLM scale on performance. Employing the same instruction data and training strategies, we fine-tune these models and present the outcomes in Fig.~\ref{modelscale}. 
The enhancements seen in LLaMA-13B in comparison to LLaMA-7B are minor. One possible explanation is
that the LLaMA-7B has already achieved strong performance.
Despite achieving better outcomes with our model utilizing LLaMA-13B, we abstain from scaling up the model due to computational costs. 
Despite having a comparable number of parameters, LLaMA-7B outperforms OPT-6.7B, thereby demonstrating the inherent strengths of LLaMA. 
Furthermore, Employing LLMs with smaller scales results in a decline in performance.
When efficiently fine-tuning a small model with 20K instruction data, the results reveal the restricted ability in knowledge retrieval owing to the insufficient scale of model parameters. Therefore, it is necessary to either employ large-scale language models or fully tune small models with more data.



\begin{figure*}[]
\centering
\includegraphics[width=2.0\columnwidth]{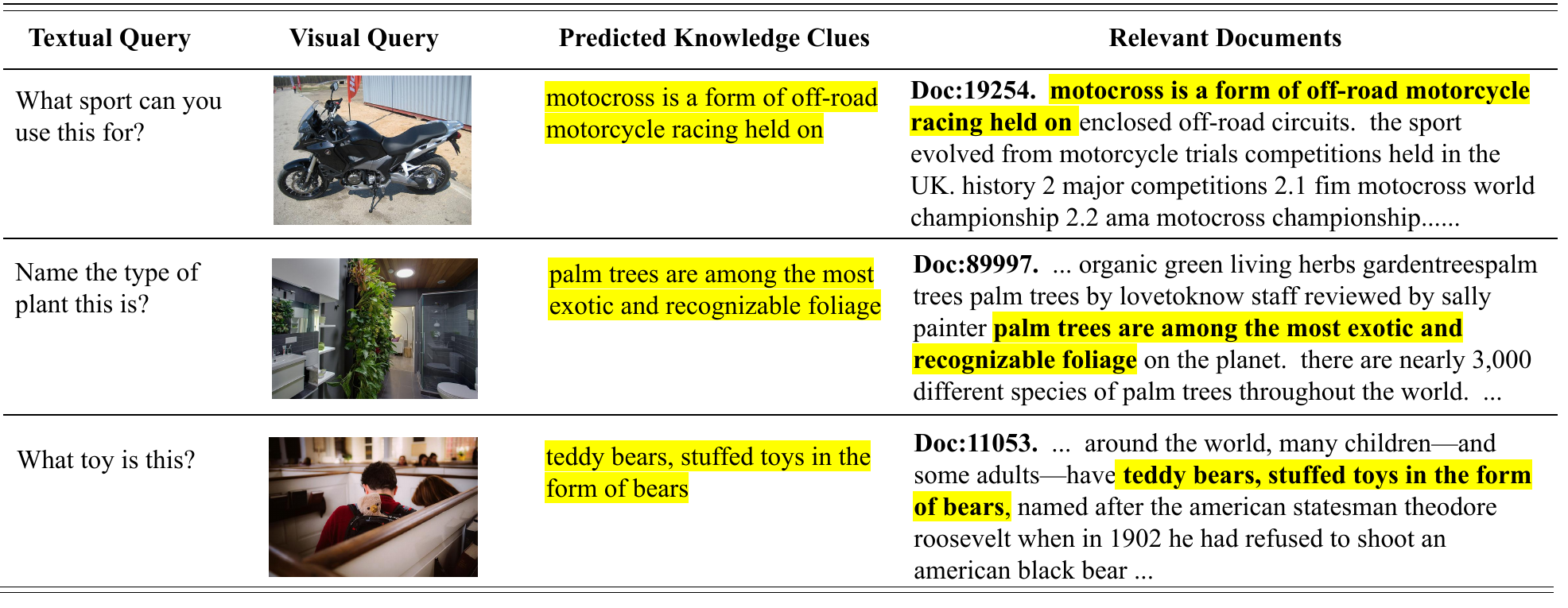}
\caption{Case Study. Three cases from the OKVQA-GS112K dataset. Each predicted knowledge clue can be uniquely mapped to a document in the KB. The predicted knowledge clues that occur in corresponding documents are highlighted in yellow.}
\label{fig12}
\end{figure*}

\subsection{Effect of Constraint Strategies}
To investigate the role of knowledge clues, we analyze the effect of constrained decoding on the recall metric. We propose four strategies with progressively relaxing constraints, 1) directly generating sentences with constraints. 2) generating the first sentence in the document under constraints, then using it as the identifier to look up the corresponding document, similar to ~\cite{10.1145/3511808.3557271}. 3) generating knowledge clues as previously described. 4) generate unconstrained text. 

\begin{table}[]
\centering
\begin{tabular}{lcccc}
\hline\hline
Generation Strategy             & R@5  & R@10 \\ \hline
Full Document w/ Constraints     & 51.6    & 59.9    \\
First Sentence w/ Constraints     & 62.4    & 70.2    \\
Knowledge Clue w/ Constraints      & 78.6 & 86.2 \\
Free Text w/o Constraints   & 64.5    & 70.9    \\
\hline\hline

\end{tabular}
\caption{Results of different generation strategies.}
\label{modeldec}
\end{table}

As shown in Tab.~\ref{modeldec}, the first and last perform poor results for two possible reasons. The one is that generating a long document from the first token is challenging due to insufficient input. The other is that strong constraints may disrupt the predicted distribution, whereas no constraint may lead to erroneous generation. 
Compared with the third, the second underperforms by at least 14\%, the possible explanation is that the
multi-modal query attends to the multiple aspects of knowledge,
while the first sentence can not represent all information in the document. 
To verify the assumption, we respectively count the occurrence of keywords in knowledge clues when using uni-modal queries as input. We sample 100 data points to visualize them in the Tab.~\ref{kcexample}

\begin{table}[t]
\centering
\begin{tabular}{cc}
\hline\hline
{\bf Only Textual Query}      & {\bf Only Visual Query}                          \\ \hline
United-States, racing,  & people, mountain,  \\
sport, American,  & building, black,  \\ 
Snowboarding, Olympics,  & white,  statue, \\
Manhattan, Boeing, & hand-painted, world, \\
Mcdonald's, Baseball &  sunrise, lightning \\  \hline\hline
\end{tabular}
\caption{Top 10 keywords in knowledge clues when using uni-modal queries.}
\label{kcexample}
\end{table}
As shown in Tab~\ref{kcexample}, the generated keywords differ across different modal inputs.
Our model tends to produce knowledge clues that align with keywords in textual queries when text alone is provided as input. Conversely, when images are the sole input, the generated knowledge clues encompass more descriptive terms related to the images, such as attributes, colors, and objects.
This observation highlights that distinct modal queries focus on diverse aspects of knowledge, indicating why the static identifier yields unsatisfactory results in this task.
Our model benefits from knowledge clues that can flexibly associate information from multiple aspects and serve as dynamic identifiers.

\subsection{Case Study}
To qualitatively illustrate why GeMKR works, we analyze the prediction results on the OKVQA-GS112K dataset in Fig.~\ref{fig12}.
We observe that (1) both textual and visual queries provide useful features. 
As seen in the third example, the term ``toy'' in the text is semantically correlated with the region depicting a ``teddy bear" in the image, 
indicating that fine-grained cross-modal correlations are important to understand multi-modal queries. 
(2)
Knowledge clues are free-format text snippets with rich semantics that can appear at any position within a document.
In contrast to static identifiers (e.g. title and Docid), knowledge clues offer greater flexibility in representing a document, harnessing the generative capabilities of LLMs without the need for additional steps to memorize associations between knowledge and identifiers.
This strategy enhances generalization for unseen knowledge, potentially contributing to the effectiveness of our model.

\section{Conclusion}
In this paper, 
we are the first to introduce a generative pipeline into multi-modal knowledge retrieval tasks, instead of discriminative retrievers, which ensemble multiple retrievers for separate modalities. 
Besides, we make use of inherent knowledge within LLMs and design an efficient fine-tuning framework to align multi-grained visual features with textual features and feed them into LLMs for efficient multi-modal learning. 
Third, we propose a novel constraint decoding strategy to utilize knowledge clues as dynamic identifiers for generative decoding.
Experiments on the three datasets demonstrate the effectiveness of our model.

\section{Acknowledgments}
This work was supported by the National Key Research and Development Program of China (2022ZD0160603).
\bibliography{aaai24}

\end{document}